\documentclass[a4paper,traditabstract]{aa}

\usepackage{natbib,amssymb,amsmath}
\usepackage{graphicx}
\usepackage{txfonts}

\begin{document}

\title{Hydrogen isotope exchanges between water and methanol \\ in interstellar ices}

\author{%
  A.~Faure      \inst{1, 2}  \and
  M.~Faure      \inst{1, 2} \and
  P.~Theul\'e   \inst{3} \and
  E.~Quirico    \inst{1, 2} \and
  B.~Schmitt    \inst{1, 2}
}

\institute{$^1$ Univ. Grenoble Alpes, IPAG, F-38000 Grenoble, France\\
$^2$ CNRS, IPAG, F-38000 Grenoble, France
  France\\\email{alexandre.faure@obs.ujf-grenoble.fr}\\ $^3$
  Aix-Marseille Universit\'e, PIIM UMR-CNRS 7345, F-13397 Marseille,
  France\\}

\date{Received / Accepted}

\titlerunning{Hydrogen isotope exchanges in interstellar ices}
\authorrunning{A. Faure et al.}

\abstract {The deuterium fractionation of gas-phase molecules in hot
  cores is believed to reflect the composition of interstellar
  ices. The deuteration of methanol is a major puzzle, however,
  because the isotopologue ratio [CH$_2$DOH]/[CH$_3$OD], which is
  predicted to be equal to 3 by standard grain chemistry models, is
  much larger ($\sim 20$) in low-mass hot corinos and significantly
  lower ($\sim 1$) in high-mass hot cores. This dichotomy in methanol
  deuteration between low-mass and massive protostars is currently not
  understood. In this study, we report a simplified rate equation model
  of the deuterium chemistry occurring in the icy mantles of
  interstellar grains. We apply this model to the chemistry of hot
  corinos and hot cores, with IRAS~16293-2422 and the Orion~KL Compact
  Ridge as prototypes, respectively. The chemistry is based on a
  statistical initial deuteration at low temperature followed by a
  warm-up phase during which thermal hydrogen/deuterium (H/D)
  exchanges occur between water and methanol. The exchange kinetics is
  incorporated using laboratory data. The [CH$_2$DOH]/[CH$_3$OD] ratio
  is found to scale inversely with the D/H ratio of water, owing to
  the H/D exchange equilibrium between the hydroxyl (-OH) functional
  groups of methanol and water. Our model is able to reproduce the
  observed [CH$_2$DOH]/[CH$_3$OD] ratios provided that the primitive
  fractionation of water ice [HDO]/[H$_2$O] is $\sim$2\% in
  IRAS~16293-2422 and $\sim 0.6$\% in Orion~KL. We conclude that the
  molecular D/H ratios measured in hot cores may not be representative
  of the original mantles because molecules with exchangeable
  deuterium atoms can equilibrate with water ice during the warm-up
  phase.
}


\maketitle

\section{Introduction}

Primordial deuterium (D) was produced during the Big Bang
nucleosynthesis with an abundance relative to hydrogen (H) of
$[$D$]$/$[$H$]_{primordial}\sim 2.8\times 10^{-5}$
\citep{fumagalli11}. Since these ``dark ages'', stellar
nucleosynthesis has converted a significant fraction of D to
$^3$He. The D/H ratio present in the protosolar nebula (PSN) 4.6~Gyr
ago is thus estimated as $[$D$]$/$[$H$]_{PSN}\sim 2.1\times 10^{-5}$
\citep{geiss98}, while the current value in the local interstellar
medium (ISM) is $[$D$]$/$[$H$]_{local}\sim 1.6\times 10^{-5}$
\citep{linsky07}. Among all the elements, the two stable isotopes of
hydrogen have the largest mass difference. The difference in
zero-point energies between a deuterated molecule and its hydrogenated
counterpart leads therefore to strong fractionation effects at low
temperature. A number of interstellar molecules are thus observed with
deuterium enhancements of several orders of magnitude, that is
molecular D/H ratios up to a few 10$^{-1}$ \citep[see the recent
  review by][]{ceccarelli14}.

Of all the various deuterated interstellar molecules, methanol is an
interesting species for several reasons. First, gas-phase methanol is
ubiquitous and solid methanol is a major constituent of the ices
coating the dust grains in the cold ISM. The fractional abundance of
gaseous methanol (relative to H) is thus $\sim 10^{-9}$ in cold clouds
and $\sim 10^{-6}$ in hot cores, where ices have
sublimated. Second, methanol is one of the most deuterated
interstellar molecules with, in particular, the triply deuterated
isotopologue identified towards the low-mass protostar IRAS~16293-2422
with a fractionation ratio $[$CD$_3$OH$]$/$[$CH$_3$OH$]$$\sim$0.01
\citep{parise04}. Third, the large abundances of methanol cannot be
explained by gas-phase reactions \citep{garrod06,geppert06}. Among
possible processes on the grain surfaces, the most efficient mechanism
in molecular clouds, seems to be the stepwise hydrogenation of solid CO
by the highly mobile H atoms \citep[see][and references
  therein]{watanabe07}. The deuteration of methanol can also proceed
through the addition of atomic deuterium on CO, or alternatively via
H/D substitution reactions on CH$_3$OH, as demonstrated experimentally
by \cite{nagaoka05,nagaoka07}. Observationally, the high deuterium
enhancements observed for CH$_2$DOH, CHD$_2$OH, and CD$_3$OH are
consistent with the statistical addition of H and D atoms to CO,
provided that the atomic $[$D$]$/$[$H$]$ ratio in the gas-phase is
$\sim$0.1-0.2 during methanol formation \citep{parise06}. In this
scenario, the deuteration of solid methanol simply scales with the
gas-phase atomic $[$D$]$/$[$H$]$ ratio. This latter is predicted to
exceed 0.1 in regions with high density and heavy depletion, such as
prestellar cores \citep{roberts03}.

Statistical models of CO hydrogenation, however, fail to explain the
low abundance of CH$_3$OD observed in low- and intermediate-mass
protostars \citep{parise06,ratajczak11}. Indeed, the gas-phase
abundance ratio [CH$_2$DOH]/[CH$_3$OD] in these sources is found to be
unexpectedly large ($\gtrsim 20$) compared to the value of 3 predicted
by grain chemistry models \citep[e.g.][]{charnley97,osamura04}. Two
main mechanisms have been suggested to explain this observational
result. First, CH$_3$OD could be selectively destroyed in the
post-evaporative gas-phase chemistry through protonation reactions
because the dissociative recombination (DR) of CH$_2$DOH$_2^+$ reforms
CH$_2$DOH, while the DR of CH$_3$ODH$^+$ can lead to both CH$_3$OD and
CH$_3$OH \citep{osamura04}. Second, laboratory experiments have shown
that H/D exchange reactions in solid state can produce a low CH$_3$OD
abundance. Thus, in the experiments of \cite{nagaoka05,nagaoka07}, H/D
substitution reactions with atomic deuterium were shown to lead to the
formation of methyl-deuterated methanol isotopologues, but no
hydroxyl-deuterated isotopologues. Thermally activated H/D exchanges
between water and methanol were also studied by \cite{souda03}, using
secondary ion mass spectrometry, and by \cite{ratajczak09}, using
Fourier-Transform Infrared (FTIR) spectroscopy. Rapid exchanges,
within a few days or less, were found to start at 120~K in the
experiment of \cite{ratajczak09} where CD$_3$OD was co-deposited in an
intimate mixture with amorphous solid water (ASW). H/D exchanges were
found to proceed through hydrogen bonds in the hydroxyl functional
groups of methanol, but not in the -CD$_3$ methyl groups, depleting
CD$_3$OD and producing CD$_3$OH and HDO. The role of ASW
crystallization in promoting molecular mobility was also emphasized by
\cite{ratajczak09}. Following this work, \cite{faure15} have recently
determined the corresponding kinetic orders and rate constants, for
both the H/D exchange and crystallization processes, in the
temperature range $120-135$~K.

The above mechanisms can thus explain, at least qualitatively, a ratio
[CH$_2$DOH]/[CH$_3$OD] much larger than 3, as observed in low- and
intermediate-mass protostars. A major puzzle, however, is that the
reverse behaviour is observed in high-mass protostars: the methanol
isotopologue ratio is found to be lower than 3,
$[$CH$_2$DOH$]$/$[$CH$_3$OD$]$ $\sim 1$, towards Orion~KL
\citep{jacq93,peng12,neill13}, W3(H$_2$O) and possibly G24.78+0.08
\citep{ratajczak11}. The similar abundances of CH$_3$OD and CH$_2$DOH
in these sources is currently not understood. A number of hypothesis
have been suggested by several authors, but as yet none is really
convincing. One possibility is that gas-phase chemistry in the hot
core phase could alter the primitive fractionation ratio set in the
ice mantles. However, in order to decrease the [CH$_2$DOH]/[CH$_3$OD]
ratio from 3 to $\sim 1$, unrealistic high [HDO]/[H$_2$O] ratios
($\gtrsim$ 0.1) are necessary \citep{charnley97,osamura04}. In
addition, the chemical timescales required to modify the primitive
fractionations are longer than the typical timescale of the hot core
phase, which is at most a few 10$^4$~yr \citep{rodgers96}. Processes
in the ice offer an alternative, but a single mechanism should
reconcile the different methanol isotopologue ratios in low- and
high-mass sources. This mechanism has been elusive so far, which led
several authors to suggest that an important piece of the puzzle is
still missing \citep{ratajczak11,peng12,neill13}.


Yet there is a robust observational result: the deuterium enrichment
of (gas-phase) organic molecules is larger in low mass than in massive
protostars. The low- and high-mass sources for which deuterium
fractionation has been best studied are IRAS~16293-2422 and Orion~KL,
respectively. Both sources were in particular the subject of recent
multi-line studies of water and methanol, including deuterated
isotopologues, using {\it Herschel}/HIFI
\citep{coutens12,coutens13,neill13}. In the Compact Ridge region
within Orion~KL, the fractionation ratio [CH$_2$DOH]/[CH$_3$OH] was
found to be $\sim 6\times 10^{-3}$ \citep{neill13} while it is
$\sim$0.4 in IRAS~16293-2422 \citep{parise06}, that is about 70 times
larger. In contrast, the [HDO]/[H$_2$O] ratio is similar in both
sources: it was estimated as $\sim 4\times 10^{-3}$ in the Compact
Ridge \citep{neill13} and it lies in the range $10^{-3}-5\times
10^{-2}$ in the hot corino of IRAS~16293-2422
\citep{coutens13,persson13,persson14}. A ratio above a few 10$^{-2}$
is however unlikely since the highest upper limit on the
$[$HDO$]$/$[$H$_2$O$]$ ratio in the ice phase is $\sim 2\times
10^{-2}$ \citep{parise03,galvez11}. As a result, the deuterium
fractionation of the (methyl-deuterated) methanol appears to be much
larger (a factor of 10 or more) than that of water in IRAS~16293-2422,
while the deuteration is similar for water and methanol in Orion~KL.

In this work, experimental kinetic data are employed in a simplified
grain surface chemistry network involving all hydrogenated and
deuterated isotopologues of water and methanol. A standard rate
equation model is employed using, in particular, the kinetic rate
constant measured by \cite{faure15} for H/D exchanges between methanol
and water in ASW. The results are compared to observations towards the
hot cores of IRAS~16293-2422 and the Orion Compact Ridge, focusing on
the evolution of the [CH$_2$DOH]/[CH$_3$OD] and [HDO]/[H$_2$O] ratios
following the sublimation of the ice mantles. The model of deuterium
chemistry is described in Sect.~2 and the results are presented in
Sect.~3. In Section~4 we discuss the implications of our results and
we conclude in Sect.~5.

\section{Model}

In the present work, a simplified grain surface chemistry network is
developed using time-dependent rate equations. The code only deals
with the hot core phase, when ices are thermally heated and
sublimated, but the post-evaporative gas-phase chemistry is
neglected. We thus assume that the hot core phase, which lasts $\sim
10^4$~yr, is too short for gas-phase chemistry to alter the deuterium
fractionation ratios set in the ice \citep[see][]{rodgers96}. Only
gas-grain interactions, i.e. thermal desorption and adsorption, and
hydrogen isotope exchange reactions in the ice are considered. We
adopt the formulation of \cite{hasegawa92} for the desorption and
accretion rates, as explained below. Molecules are physisorbed onto
standard dust grains, which are characterized by a radius of 0.1$\mu
m$, a density of 3g.cm$^{-3}$ and $1.3\times 10^6$ surface sites for
adsorption (i.e. one monolayer), corresponding to a surface density of
sites $n_s=1.0\times 10^{15}$~cm$^{-2}$. With a gas-to-dust ratio of
100 by mass, the number density of the grains, $n_d$, is $1.3\times
10^{-12}n_{\rm H}$, where $n_{\rm H}$ is the total abundance of
hydrogen atoms, i.e.  $n_{\rm H}=n({\rm H}) + 2n({\rm H_2})$. With the
ice compositions given below, the total number of monolayers on a
grain is $\sim$30. In all calculations, we assume that the gas and
dust temperatures, $T_g$ and $T_d$, are equal.

\subsection{Thermal desorption and accretion} 

The desorption energies $E_d$ of all species are that of water, taken
from \cite{sandford88}. It is thus assumed that all species co-desorb
with water, as found experimentally for CH$_3$OH co-deposited in an
intimate mixture with water \citep{collings04}. In addition,
desorption follows a first-order kinetics, as presumed by
\cite{sandford88}. In fact, many recent studies on the thermal
desorption of thick multi-layer ices have shown that the desorption
kinetics resemble zeroth-order rather than first order reaction
kinetics \citep{fraser01,brown07,burke10}. We have checked, however,
that the half-lives of H$_2$O ice derived from these studies
\citep[e.g.][]{brown07} are in good agreement with that of
\cite{sandford88}. The desorption rate was thus taken as
\begin{equation}
k_{des}=\nu_{des}\exp\left(\frac{-E_d}{T}\right),
\end{equation}
with $\nu_{des}=2.0\times 10^{12}$~s$^{-1}$ and $E_d=5070$~K for
annealed ASW \citep{sandford88}. At 100~K, this rate is $\sim
1.9\times 10^{-10}$~s$^{-1}$, corresponding to a half-life of
116~yr. This can be compared with the value of 67~yr deduced from the
experimental data of \cite{brown07}. Desorption by cosmic rays and
photodesorption are neglected in our model.

The sticking coefficient for all species striking a grain is assumed
to be 1.0 \citep[see e.g.][for H$_2$O sticking on H$_2$O
  ice]{gibson11}. The accretion rate for a given molecule $i$ is
therefore \citep{hasegawa92}:
\begin{equation}
k_{acc}(i)=\sigma_d\langle v(i)\rangle n_d,
\end{equation}
where $\sigma_d$ is the geometrical dust-grain cross section and
$\langle v(i)\rangle$ is the thermal velocity of molecule $i$. At
100~K and for $n_{\rm H}=2\times 10^7$~cm$^{-3}$, this rate is $\sim
2.8\times 10^{-10}$~s$^{-1}$ for H$_2$O. Grains are neutral in our
model.

\subsection{Hydrogen isotope exchanges}

In contrast to usual grain chemistry models, here the adsorbed species
do not require large scale mobility to react. This is, first, because
the reactant H$_2$O is in large excess and, as a result, each
deuterated methanol molecule is always surrounded by four H$_2$O
molecules. Second, H/D exchanges only require protons and deuterons to
be labile in the ice. We therefore do not need to include the
diffusion of molecular species via thermal hopping or quantum
tunnelling in the rate equations. On the other hand, the lability of
protons and deuterons, as well as the mobility of water molecules
triggerred by ASW crystallization, is included through the
experimental H/D exchange rate constant.

The experiment of \cite{ratajczak09} had demonstrated that H/D
exchanges between methanol and water occur in the hydroxyl functional
group of methanol, but not in the methyl group. In the study of
\cite{faure15}, the kinetics of the following reaction was thus
measured:
\begin{equation}
\rm CD_3OD + H_2O \rightleftharpoons CD_3OH + HDO,
\label{reac1}
\end{equation}
by monitoring the time evolution of HDO with FTIR spectroscopy. The
rate constant for the forwards direction was shown to follow a simple
Arrhenius law over the temperature range 120-135~K, 
\begin{equation}
k_{exch}=\nu_{exch}\exp\left(\frac{-E_{exch}}{T}\right),
\label{kexch}
\end{equation}
with $\nu_{exch}=9.7\times 10^{9}$~s$^{-1}$ and $E_{exch}=4100$~K if a
first-order kinetics is assumed. The uncertainty on the activation
energy $E_{exch}$ is about 900~K, which translate into three orders of
magnitude uncertainties in the pre-exponential factor $\nu_{exch}$
\citep{faure15}. It is therefore crucial to adopt the previous
numerical values of $\nu_{exch}$ and $E_{exch}$ as a kinetic
doublet. In fact, the experimental data of \cite{faure15} are best
reproduced by a second- or third-order kinetic model. A first-order
approximation is adopted here for simplicity, but with the
corresponding kinetic doublet. Finally, it should be noted that
$k_{exch}$ is an apparent or pseudo first-order rate constant since in
the experiment H$_2$O was in great excess compared to CD$_3$OD and
thus $k_{exch}$ does not depend on the order of the reaction with
respect to H$_2$O (i.e. the partial order of H$_2$O). At 100~K, the
above pseudo first-order rate is $\sim 1.5\times 10^{-8}$~s$^{-1}$,
corresponding to a half-life of 1.5~yr. This timescale is thus
significantly shorter than the timescales for thermal desorption and
accretion.

In the standard rate equations described below, the reaction flux
$R_{i; j}$, in cm$^{-3}$s$^{-1}$, between two adsorbed species $i$ and
$j$ is expressed in terms of a rate constant $k_{i; j}$, in
cm$^3$s$^{-1}$, analoguous to those employed in gas-phase two-body
reactions
\begin{equation}
R_{i; j}=k_{i; j}n_s(i)n_s(j),
\end{equation}
where $n_s(i)$ is the surface abundance of species $i$. The above
first-order rate constant $k_{exch}$ is thus related to the
second-order rate constant $k_{\rm CD_3OD; H_2O}$ {\it via} the
  expression
\begin{equation}
k_{exch}=k_{\rm CD_3OD; H_2O}n_s(\rm{H_2O}),
\end{equation}
assuming a H$_2$O partial order of 1. The two-body rate constant for
H/D exchanges between CD$_3$OD and H$_2$O is, therefore, 
\begin{equation}
k_{\rm{CD_3OD; H_2O}}=k_{exch}/n_s(\rm{H_2O}).
\label{kh2o}
\end{equation}
Thus, the reaction rate ``constant'' for H/D exchanges vary with time
depending on the surface abundance of H$_2$O. This variation is
however weak in practice (about 40\% at 100~K) and it is mainly caused
by sublimation. This treatment is analoguous to that of the grain
surface formation of H$_2$ in gas-phase networks where the rate
constant depends on the gas-phase abundance of hydrogen atoms.

As a consequence of the above treatment, the exchange rate constants
$k_{i; j}$ were set to zero in two circumstances: {\it i)} if the
abundance of a solid reactant is below that of grains (unphysical); and
{\it ii)} if the number of monolayers is lower than one. This latter
condition was however found to occur only for grain temperatures above
110~K.

\subsection{Reversibility}

H/D exchange reactions are almost thermoneutral and they are
reversible in the ice \citep{collier84}. This is exemplified in the
chemical Eq.~(\ref{reac1}) above. In the experiment of
\cite{faure15}, only the forwards reaction rate constant was measured
but microscopic reversibility predicts a relationship between the
forwards $k_f$ and backwards $k_b$ rate constants: their ratio,
$k_f/k_b$, should be equal to $K(T)$, the equilibrium constant. In the
theory for gaseous molecules, this constant depends on the masses and
partition functions of the reactants and products, and on the
difference between their zero-point vibrational energies
\citep[e.g.][]{richet77}. A simplified expression for $K(T)$ in liquid
solutions is \citep{bertrand75,fenby82} as follows:
\begin{equation}
K(T)=K^{stat}\exp(-\Delta H/T),
\label{kstat}
\end{equation}
where $\Delta H$ is the effective enthalpy of the reaction and
$K^{stat}$ is the (high temperature) statistical equilibrium constant
based on a completely random distribution of exchangeable (identical)
particles such as protons and deuterons. We stress that this equation
requires that the (standard) entropy of reaction is equal to
$\ln(K^{stat})$ at all temperature.

In the case of the prototypical H/D exchange reaction
\begin{equation}
\rm D_2O + H_2O \rightleftharpoons HDO + HDO,
\label{reac2}
\end{equation}
it can be easily shown that $K^{stat}=4$. The enthalpy $\Delta H$ is
unknown in the ice phase, but the equilibrium constant has been
measured in the liquid and gas phase. Although there is some
variation in published values, a reference experimental value for the
liquid phase is $K$(298~K)=3.82 \citep{simonson90}, which compares
well with the gas-phase theoretical value\footnote{A theoretical value
  can be obtained from the {\it ab initio} data published in
  \cite{hewitt05}.} $K$(300~K)=3.83. This equilibrium constant
$K$(298~K)=3.82 corresponds to an effective enthalpy $\Delta H=14$~K,
which is employed below.

Similarly, in the case of the reaction
\begin{equation}
\rm CH_3OD + H_2O \rightleftharpoons CH_3OH + HDO,
\label{reac3}
\end{equation}
one can show that $K^{stat}=2$. The equilibrium constant for this
reaction was also measured in liquid phase by several authors
\citep[see][and references therein]{zhang92}. Values at 298~K are
dispersed in the range $1.4-2$ and we have thus adopted $\Delta
H=50$~K in the following.

We stress that the notion of reversibility may seem contradictory with
the irreversible behaviour of the crystallization process, which
promotes H/D exchanges. As discussed in \cite{faure15}, however, H/D
exchanges occur in the {\it \textup{very initial} \textup{phase}} of
crystallization, during the nucleation process. The reversibility of
H/D exchanges is thus consistent with the idea that amorphous water
ice melts into a metastable liquid-like state prior to crystallizing
(the glass transition), as suggested experimentally \cite[see
  e.g.][]{smith99}. In other words, crystallization is not expected to
prevent exchanges to reach equilibrium.

In addition to reaction~(\ref{reac3}), seven other reversible
reactions between the hydroxyl-deuterated isotopologues of methanol
and water are considered:
\begin{equation}
\rm CH_2DOD + H_2O \rightleftharpoons CH_2DOH + HDO,
\end{equation}
\begin{equation}
\rm CHD_2OD + H_2O \rightleftharpoons CHD_2OH + HDO,
\end{equation}
\begin{equation}
\rm CD_3OD + H_2O \rightleftharpoons CD_3OH + HDO,
\label{reac4}
\end{equation}
\begin{equation}
\rm CH_3OD + HDO \rightleftharpoons CH_3OH + D_2O,
\label{reac5}
\end{equation}
\begin{equation}
\rm CH_2DOD + HDO \rightleftharpoons CH_2DOH + D_2O,
\end{equation}
\begin{equation}
\rm CHD_2OD + HDO \rightleftharpoons CHD_2OH + D_2O,
\end{equation}
\begin{equation}
\rm CD_3OD + HDO \rightleftharpoons CD_3OH + D_2O.
\label{reac6}
\end{equation}
Reactions~(\ref{reac3})-(\ref{reac4})  all proceed with the
forwards second-order rate constant
\begin{equation}
k_f=9.7\times 10^{9}\exp(-4100/T){\rm s^{-1}}/n_s(\rm{H_2O}),
\label{kf}
\end{equation}
as deduced from Eqs.~(\ref{kexch})-(\ref{kh2o}), and the backwards
(second-order) rate constant
\begin{equation}
k_b=\frac{1}{2}k_f\exp(50.0/T),
\label{kb}
\end{equation} 
as deduced from the measurements in liquid water
\citep{bertrand75,zhang92}. We thus assume that all
reactions~(\ref{reac3})-(\ref{reac4}) have the same enthalpy $\Delta
H=50$~K. Similarly, reactions~(\ref{reac5})-(\ref{reac6})  proceed
with a forwards rate constant
\begin{equation}
k_f=9.7\times 10^{9}\exp(-4100/T){\rm s^{-1}}/n_s(\rm{HDO}),
\label{kf2}
\end{equation}
and a backwards rate constant
\begin{equation}
k_b=2k_f\exp(36.0/T).
\label{kb2}
\end{equation} 
Here we assume that H/D exchanges with HDO (which are of secondary
importance in our network because of the predominance of H$_2$O) proceed
with the same (first-order) forwards rate constant as with H$_2$O. The
enthalpy of these reactions is taken as $\Delta H=36$~K, which is the
enthalpy difference between reaction~(\ref{reac3}) and
reaction~(\ref{reac2}). Indeed, in case of multiple equilibria, the
principle of addition of enthalpies (Hess's law) allows us to calculate
unknown enthalpy change from the combination of reactions.

The last reversible reaction is given by Eq.~(\ref{reac2}) for
which the forwards rate constant is taken as in Eq.~(\ref{kf}) and
the backwards rate constant as
\begin{equation}
k_b=\frac{1}{4}k_f\exp(14.0/T).
\label{kb3}
\end{equation}
It is thus assumed that H/D exchanges in D$_2$O + H$_2$O proceed with
the same rate constant as in CH$_3$OD + H$_2$O. This hypothesis is
consistent with a number of measurements in water ice
\citep{park10,lamberts15}. In particular, the very recent FTIR
experiment of \cite{lamberts15} on reaction~(\ref{reac2}) in ASW and
for temperatures in the 90-140~K domain, has provided an activation
energy of $3840\pm 125$~K, which is in good agreement with the value
of $4100\pm 900$~K determined by \cite{faure15} for
reaction~(\ref{reac1}). Reaction~(\ref{reac2}) was also studied in the
range 135-150~K in cubic ice by \cite{collier84}, who found an
activation energy of $\sim 4800$~K. In the study of \cite{collier84},
the H/D exchange mechanism was separated into a two-step reaction
sequence, the so-called hop-and-turn mechanism. This study was
extended to ASW samples in the 108 to 125~K domain by \cite{fisher95}
and a significantly lower activation energy of $\sim$3000~K was
measured. However, protons were injected into the ASW samples in this
case. Finally, \cite{park04} have also measured H/D exchanges between
water molecule at ice surfaces and they obtained a rate of $\sim
6\times 10^{-4}$~s$^{-1}$ at 140~K in rather good agreement with the
value of $2\times 10^{-3}$~s$^{-1}$ obtained from Eq.~(\ref{kexch})
above. However, as a result of the very low abundance of D$_2$O with
respect to HDO and CH$_3$OD in our model, the reaction between D$_2$O
and H$_2$O is of minor importance in our methanol network.

Finally, we emphasize that the kinetic parameters of
equation~(\ref{kexch}) were derived experimentally by \cite{faure15}
over the limited temperature range 120-135~K. They are expected to
apply over a much larger range of temperatures, although deviations
from the Arrhenius behaviour are likely at very low temperatures due to
quantum effects \citep[e.g.][]{mispelaer12}.

\subsection{Surface reaction network}

The ice-phase species considered above are coupled to the gas phase
through the accretion and thermal sublimation processes. The rate
equations for a species $i$ in the ice phase with concentration
$n_s(i)$ and in the gas phase with concentration $n(i)$ are given by
the following expressions:
\begin{eqnarray}
dn_s(i)/dt  & =  & \sum_l\sum_j k_{l; j}n_s(l)n_s(j) - n_s(i)\sum_j k_{i; j}n_s(j) \nonumber \\ 
            &    & + n(i)k_{acc}(i) - k_{evap}n_s(i), \label{tsurf}   \\
dn(i)/dt    & =  &  -n(i)k_{acc}(i) + k_{evap}n_s(i). \label{tgas}
\end{eqnarray}
The above equations are identical to standard gas-grain chemical
models \citep[e.g.][]{hasegawa92} except that {\it i)} diffusion of
adsorbed species is not needed and therefore purposely not included in
Eq.~(\ref{tsurf}); and {\it ii)} gas-phase reactions are ignored in
Eq.~(\ref{tgas}). Chemical reactions are thus restricted to H/D
exchanges in the solid phase through the rate constants $k_{i; j}$ defined
in Eqs.~(\ref{kf})-(\ref{kb3}).

The numerical method employed to solve the previous coupled
differential equations is adapted to stiff ordinary differential
equations (ODE) systems, as implemented in the \texttt{LSODE} solver
from the \texttt{ODEPACK} package \citep{hindmarsh83}. Abundances are
followed up to steady-state, which occurs within $\sim 10^3$~yr at a
density of $n_{\rm H}=2\times 10^7$~cm$^{-3}$ and $T$=100~K. There are
nine reversible surface reactions in our network,
Eq.~(\ref{reac2})-(\ref{reac6}), involving eleven species listed in
Tables~1-2 below.

\subsection{Initial conditions}

The chemical model was run for two different initial conditions,
representative of the ice composition measured towards low-mass and
massive protostars. The initial ice-phase abundances were thus taken
as the median ice composition determined by \cite{oberg11}: the
methanol abundance (by number) relative to water is 3~\% towards
low-mass protostars and 4~\% towards high-mass protostars. A mean
H$_2$O ice abundance of $5\times 10^{-5}n_{\rm H}$,
i.e. $10^{-4}n({\rm H_2})$, is assumed for both low- and high-mass
sources \citep{oberg11}. The initial fractionation ratios were
computed using a purely statistical model. Thus, if $p_{\rm D}$ is the
probability of accretion of D atoms on a grain, the statistical
fractionation ratio of HDO ($R({\rm HDO})\equiv [{\rm HDO}]/[{\rm
    H_2O}]$) is
\begin{eqnarray}
\label{stat1}
R({\rm HDO}) & = & 2p_{\rm D}/(1-p_{\rm D}) = 2\alpha_w,
\end{eqnarray}
where $\alpha_w$ is the accreting D/H ratio during the formation of
water, $\alpha_w=[\rm D]/[\rm H]/\sqrt{2}$. The $\sqrt{2}$ arises
because the accretion of an atom is inversely proportional to the
square root of its mass (assuming the sticking coefficient is
identical for H and D). Similarly, for D$_2$O, we find 
\begin{eqnarray}
\label{stat2}
R({\rm D_2O}) & = & \alpha_w^2.
\end{eqnarray}
As discussed below, the accreting D/H ratio can vary with time and we
have therefore assumed that methanol can form with an accreting D/H
ratio $\alpha_m$ different from that of water. We find therefore for
the isotopologues of methanol as follows:
\begin{eqnarray}
\label{stat3}
R({\rm CH_3OD}) & = & \alpha_m ,\\
\label{stat4}
R({\rm CH_2DOH}) & = & 3\alpha_m ,\\
R({\rm CH_2DOD}) & = & 3\alpha_m^2 ,\\
R({\rm CD_2HOH}) & = & 3\alpha_m^2 ,\\
R({\rm CD_2HOD}) & = & 3\alpha_m^3 ,\\
R({\rm CD_3OH}) & = & \alpha_m^3 ,\\
\label{stat9}
R({\rm CD_3OD}) & = & \alpha_m^4
.\end{eqnarray}
The $\alpha_w$ and $\alpha_m$ ratios cannot be directly measured in
the interstellar gas. The $\alpha_m$ ratio can, however, be inferred
from the measured D/H ratios of gaseous methanol, assuming that the
initial (statistical) deuteration of CH$_2$DOH in the ice is conserved
during the heating and hot core phase \citep{ratajczak09,faure15}. In
this hypothesis, the observed fractionation ratio $R({\rm CH_2DOH})$
can serve as a proxy of $\alpha_m$. As discussed in the Introduction,
the best studied low- and high-mass protostars are IRAS~16293-2422 and
Orion~KL, respectively. In IRAS~16293-2422, the gas-phase $R({\rm
  CH_2DOH})$ ratio was found to be $\sim$0.37
\citep{parise06}\footnote{Slightly larger values were
  derived for the other low-mass sources IRA4A, IRAS4B and IRAS2
  \citep{parise06}.}. In the Orion Compact Ridge, this ratio is $\sim
5.8\times 10^{-3}$ \citep{neill13}. From Eq.~(\ref{stat4}), these
values correspond to $\alpha_m$ ratios of $\sim$0.12 and $\sim 2\times
10^{-3}$. These values were employed in
Eqs.~(\ref{stat3})-(\ref{stat9}) to specify the initial conditions
listed in Table~1 for IRAS-16293-2422 and in Table~2 for the Compact
Ridge. The different values of $\alpha_m$ in low- and high-mass
protostars are discussed in Section~4. In the case of water,
Eqs.~(\ref{stat1})-(\ref{stat2}) were used but a large range of
$\alpha_w$ values between $10^{-5}$ and $5\times 10^{-2}$ was adopted,
based on the observational constraints. The D/H ratio of water ice is
thus employed as a free parameter in our model, in agreement with the
hypothesis that the deuterium fractionation ratios of solid water and
methanol can be different.

\begin{table*}[htp]
\centering
\caption{Initial conditions for the ice composition in
  IRAS~16293-2422. The fractionation ratios of the deuterated
  isotopologues of methanol are purely statistical assuming an
  accreting ratio $\alpha_m$ of 0.12. Numbers in brackets denote the
  power of ten, e.g. $1.0(-4)=1.0\times 10^{-4}.$}
\begin{tabular}{c c c c}
\hline
\hline
Species    & Abundance  & Isotopologue & Fractionation  \\
\hline
H$_2$O     & 5.0(-5)    & HDO          & 1.0(-4)$-$1.1(-1)    \\ 
           &            & D$_2$O       & 2.5(-9)$-$2.8(-3)  \\ 
CH$_3$OH   & 1.5(-6)    & CH$_3$OD     & 1.2(-1)   \\ 
           &            & CH$_2$DOH    & 3.6(-1)  \\ 
           &            & CH$_2$DOD    & 4.3(-2)  \\ 
           &            & CD$_2$HOH    & 4.3(-2)  \\ 
           &            & CD$_2$HOD    & 5.2(-3)  \\ 
           &            & CD$_3$OH     & 1.7(-3)  \\ 
           &            & CD$_3$OD     & 2.1(-4)  \\ 
\hline
\end{tabular}
\end{table*}

\begin{table*}[htp]
\centering
\caption{Initial conditions for the ice composition in Orion~KL. The
  fractionation ratios of the deuterated isotopologues of methanol are
  purely statistical assuming an accreting ratio $\alpha_m$ of
  $2\times 10^{-3}$.}
\begin{tabular}{l c l c}
\hline
\hline
Species    & Abundance  & Isotopologue & Fractionation  \\
\hline
H$_2$O     & 5.0(-5)    & HDO          & 1.0(-4)$-$1.1(-1)    \\ 
           &            & D$_2$O       & 2.5(-9)$-$2.8(-3)  \\ 
CH$_3$OH   & 2.0(-6)    & CH$_3$OD     & 2.0(-3)   \\ 
           &            & CH$_2$DOH    & 6.0(-3)  \\ 
           &            & CH$_2$DOD    & 1.2(-5)  \\ 
           &            & CD$_2$HOH    & 1.2(-5)  \\ 
           &            & CD$_2$HOD    & 2.4(-8)  \\ 
           &            & CD$_3$OH     & 8.0(-9)  \\ 
           &            & CD$_3$OD     & 1.6(-11)  \\ 
\hline
\end{tabular}
\end{table*}


Finally, elaborate theoretical models of the grain surface chemistry
of deuterium exist in the literature
\cite[e.g.][]{charnley97,cazaux11,sipila13,taquet14,lee15}. These are
partly based on laboratory data, which may suggest deviations from
statistical distributions in some cases. Modelling the deuterium
fractionation processes of ices is clearly beyond the scope of the
present paper, which intentionally focuses on a unique surface
chemistry process, the deuteron scrambling in water ice.

\section{Results}

The coupled set of rate equations~(\ref{tsurf})-(\ref{tgas}) was
solved subject to the initial conditions listed in Tables~1 and 2 for
IRAS~16293-2422 and the Orion Compact Ridge, respectively. The density
was fixed at $n_{\rm H}=2\times 10^7$~cm$^{-3}$ for both sources,
which is typical of hot core regions. The (equal) gas and dust
temperatures, $T_g$ and $T_d$, were varied between 60 and 140~K,
assuming an instantaneous temperature jump. For grain temperatures
above 90~K, the steady-state was reached in less than $10^3$~yr, which
is shorter than the duration of the hot core phase. Only the
steady-state solutions are  therefore discussed in the following.

\subsection{Low-mass protostar: IRAS~16293-2422}

\begin{figure}
\begin{center}
\includegraphics*[width=7.0cm,angle=-90.]{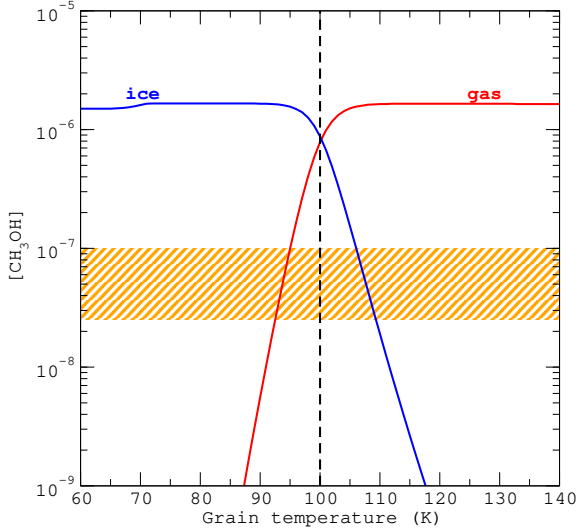}
  \caption{Gas phase and ice abundance of CH$_3$OH as a function of
    grain temperature. The abundance observed towards IRAS~16293-2422
    is represented by the orange hatched zone \citep{parise04}. The
    solid vertical dashed line denotes the standard hot-core dust
    temperature $T_d=100$~K. See text for details.}
\label{fig1}
\end{center}
\end{figure}

The steady-state gas- and ice-phase abundances of CH$_3$OH in
IRAS~16293-2422 are plotted in Fig.~\ref{fig1} as a function of dust
temperature. In this plot, the initial solid water fractionation ratio
is taken as [s-HDO]/[s-H$_2$O]=1\%. We can observe that for
temperatures lower than 85~K, the gas-phase abundance of CH$_3$OH
(relative to $n_{\rm H}$) is almost negligible, i.e. below
10$^{-9}$. However, above this critical temperature, the abundance
increases sharply and solid methanol is entirely sublimated for
temperatures higher than 110~K. We find the high
temperature gas-phase abundance ($1.6\times 10^{-6}$) is slightly
larger than the low-temperature ice-phase abundance ($1.5\times
10^{-6}$). As we  show below, this difference is caused by the H/D
exchange reaction between methanol and water.

The CH$_3$OH abundance (relative to $n_{\rm H}$) in IRAS~16293-2422
was estimated as $\sim 5\times 10^{-8}$ using a CH$_3$OH column
density of $\sim 10^{16}$~cm$^{-2}$, as derived by \cite{parise04},
and a H$_2$ column density of $\sim 10^{23}$~cm$^{-2}$ for the hot
corino of IRAS~16293-2422 \citep{chandler05}. The orange hatched zone
in Fig.~\ref{fig1} depicts the corresponding uncertainty of a factor
of 2. It is found from Fig.~\ref{fig1} that the observed abundance is
consistent with a dust temperature of $\sim 95$~K. Hence, only a small
fraction of the ice mantles seem to have evaporated into the gas phase
(assuming that the ice composition in IRAS~16293-2422 is typical of
low-mass sources). We recall, however, that in addition to an
instantaneous temperature jump our model assumes single-size and
single-temperature dust grains. If grains with different sizes have
different temperatures, then ice mantles could be only partially
released into the gas phase because of the grain temperature distribution
\citep[see e.g.][]{maret05}. In the following, we  therefore
employ a standard hot-core temperature $T_d=$100~K. This choice
implies that warm ($T_d\geq 100$~K) mantles represent only a few
percent of the total ice reservoir in IRAS~16293-2422. We have also
checked that our conclusions are unchanged when using $T_d=95$~K.

\begin{figure}
\begin{center}
\includegraphics*[width=7.0cm,angle=-90.]{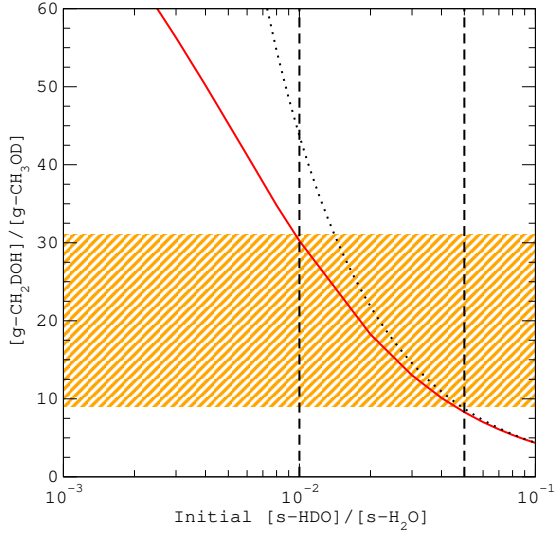}
  \caption{Gas-phase abundance ratio of the singly deuterated
    isotopologues of methanol, [g-CH$_2$DOH]/[g-CH$_3$OD], as function
    of the initial water ice deuteration. The dust temperature is
    taken as 100~K. The dotted line corresponds to
    Eq.~(\ref{equi3}). The ratio observed towards IRAS~16293-2422 is
    represented by the orange hatched zone \citep{parise04}.}
\label{fig2}
\end{center}
\end{figure}

The gas-phase isotopologue ratio [g-CH$_2$DOH]/[g-CH$_3$OD] obtained
with our model is plotted in Fig.~\ref{fig2} as function of the free
parameter, the initial water ice deuteration [s-HDO]/[s-H$_2$O]. The
ratio is found to exceed the statistical value of 3 over the entire
range of initial water deuteration. Isotopic H/D exchanges between
methanol and water at $\sim$100~K can thus  {\it
  \textup{quantitatively explain}} the large [g-CH$_2$DOH]/[g-CH$_3$OD] ratios observed
in low-mass protostars. In particular, our model matches the
observational ratio towards IRAS~16293-2422 ($20\pm 11$) for an
initial [s-HDO]/[s-H$_2$O] in the range $\sim 1-5\times 10^{-2}$. This
range of values is consistent with the observational upper limits
derived towards low-mass protostars, i.e. [s-HDO]/[s-H$_2$O]$\leq
2\times 10^{-2}$ \citep{parise03,galvez11}, and with the most recent
theoretical works, i.e. [s-HDO]/[s-H$_2$O]$\leq 5\times 10^{-2}$
\citep{lee15}. The observation of the two singly deuterated forms of
methanol in the gas-phase thus provides an indirect measurement of the
primitive deuteration of water, which has sublimated in the hot corino
region. The large difference between this primitive D/H ratio
([s-HDO]/[s-H$_2$O]$\sim 2\%$) and that of methanol
([s-CH$_2$DOH]/[s-CH$_3$OH]$=36\%$, see Table~1) implies a strong
heterogeneity of the molecular D/H enrichments in the grain
mantles. This is further discussed in the next section.

We can now try to understand the above result analytically. The
increase of the [g-CH$_2$DOH]/[g-CH$_3$OD] ratio with decreasing water
deuteration was in fact expected since the amount of gaseous CH$_3$OD
should scale with [s-HDO]/[s-H$_2$O]. Indeed, since the timescale for
isotopic exchanges at 100~K is short compared to the desorption and
accretion timescales (see Sections~2.1 and 2.2), all H/D exchange
reactions~(\ref{reac2})-(\ref{reac6}) attain equilibrium in the
ice. In the case of the dominant process, Eq.~(\ref{reac3}), we can
write at steady-state and at chemical equilibrium
\begin{equation}
\frac{[{\rm CH_3OH}][{\rm HDO]}}{[{\rm CH_3OD}][{\rm
      H_2O}]}=\frac{k_f}{k_b}=2\exp(-50/T_d)
\label{equi}
.\end{equation}
The above equation holds for both the ice and the gas phase, since
the desorption rate is taken equal for all species. This was checked
and confirmed numerically. As the initial abundance of CH$_2$DOH is
conserved in the ice (since no H/D exchange occur on the methyl
group), we can combine Eq.~(\ref{stat4}) and Eq.~(\ref{equi}) and we
obtain
\begin{equation}
\frac{[{\rm CH_2DOH}]}{[{\rm
      CH_3OD}]}=3\alpha_m\times 2\exp(-50/T_d)\frac{[{\rm
      H_2O}]}{[{\rm HDO}]}
\label{equi2}
.\end{equation}
For a dust temperature $T_d=100$~K and $\alpha_m=0.12$, we finally get
\begin{equation}
\frac{[{\rm CH_2DOH}]}{[{\rm CH_3OD}]}=\frac{0.44}{[{\rm HDO}]/[{\rm
      H_2O}]}
\label{equi3}
.\end{equation}
Equation~(\ref{equi3}) is plotted in Fig.~\ref{fig2} as a dotted line. We
can observe that it reproduces very well the ratio
[g-CH$_2$DOH]/[g-CH$_3$OD] for initial water deuteration larger than
$\sim 5\times 10^{-2}$. In this regime, the main reservoir of {\it
  \textup{exchangeable}} deuterium atoms in the ice is water
([s-HDO]$>2.5\times 10^{-6}$) and the HDO abundance is not modified by
H/D exchanges with methanol since the (initial) CH$_3$OD abundance is
only $1.8\times 10^{-7}$. In contrast, at lower initial water
deuteration, i.e. [s-HDO]/[s-H$_2$O]$\lesssim 5\times 10^{-2}$, the
relative amount of deuterium atoms in CH$_3$OD becomes significant. In
this regime, water gets enriched in deuterium through H/D exchanges
with CH$_3$OD and the model deviates from Eq.~(\ref{equi3}) since the
solid HDO abundance has increased.

\begin{figure}
\begin{center}
\includegraphics*[width=7.0cm,angle=-90.]{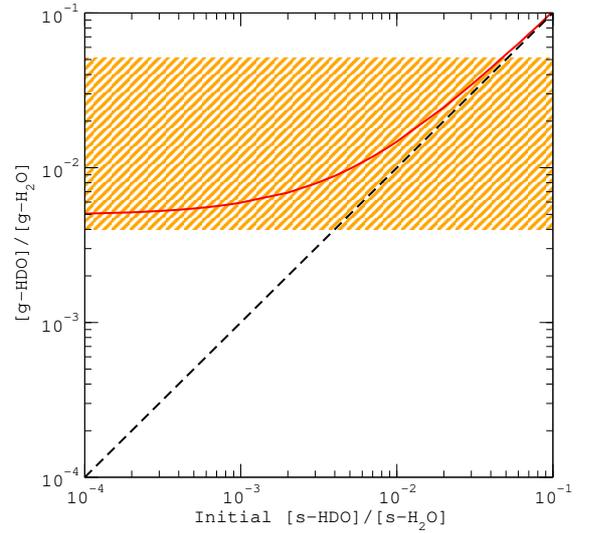}
  \caption{Gas-phase deuterium fractionation of water,
    [g-HDO]/[g-H$_2$O], as function of the initial water ice
    deuteration. The ratio observed towards IRAS~16293-2422 is
    represented by the orange hatched zone \citep{coutens13}.}
\label{fig3}
\end{center}
\end{figure}

In order to illustrate this last point, in
Fig.~\ref{fig3} we plotted  the D/H ratio of gaseous water as function of the
initial D/H ratio of water ice. If the D/H ratio of water ice was
conserved during the sublimation process, i.e. if there was no H/D
exchange, then the curve would follow the $y=x$ function represented
by the dashed line. As expected, however, when the initial
[s-HDO]/[s-H$_2$O] is lower than $\sim 5\times 10^{-2}$, the gas-phase
ratio is systematically increased with respect to the initial
ice-phase value, by up to a factor of 50 at
[s-HDO]/[s-H$_2$O]=$10^{-4}$. For the best value
[s-HDO]/[s-H$_2$O]=2\% (see Fig.~\ref{fig2}), the [g-HDO]/[g-H$_2$O]
values is found to be $2.5\times 10^{-2}$. As shown in
Fig.~\ref{fig3}, this is consistent with the range of observational
ratios [g-HDO]/[g-H$_2$O] measured in the hot corino of IRAS~16293 by
\cite{coutens13}. We emphasize that observational uncertainties are large
owing to optical thickness problems. In any case, if the
post-evaporative chemistry has no significant effect on the deuterium
and water chemistry, as assumed here, our model then predicts a D/H
ratio for gaseous water of $\sim 2.5\%$ in the hot corino of
IRAS~16293-2422. This value is in good agreement with the
best-fit value of 1.8~\% derived by \cite{coutens13} from their
multi-line analysis. On the other hand, if the low value of $9.2\pm
2.6 \times 10^{-4}$ derived from the interferometric observations of
\cite{persson13} is confirmed, then our model suggests that a
significant additional production of gaseous H$_2$O must occur during
the (short) hot core phase.

Finally, in Table~3, we summarize the results of our best model,
corresponding to an initial [s-HDO]/[s-H$_2$O] of 2\%. A good
agreement with the observations is obtained, including the multiply
deuterated isotopologues. The next isotopologue to be discovered is
obviously CH$_2$DOD, for which our model predicts a column density of
$\sim 1.2\times 10^{15}$~cm$^{-2}$ or an abundance (relative to
$n_{\rm H}$) of $5.8\times 10^{-9}$. To the best of our knowledge,
however, experimental frequencies in the millimetre range are not
available for this isotopologue. We can also notice that the
fractionation ratio of D$_2$O predicted by our model ($1.7\times
10^{-4}$) agrees within a factor of $\sim$2 with the best-fit value of
$7\times 10^{-5}$ derived by \cite{coutens13} The value quoted in
Table~3 is the upper limit at 3$\sigma$.

\begin{table*}[htp]
\centering
\caption{Comparison between observational fractionation ratios in the
  IRAS~16293-2422 hot corino source and our best model. Numbers in
  brackets denote the power of ten.}
\begin{tabular}{l c c c}
\hline
\hline
Species    & Best model  & Observations                &  References  \\ [+3pt]
\hline
HDO        &   2.5(-2)   &  6.6(-4)-5.0(-2)            & \cite{coutens13} \\
           &             &                             & \cite{persson14} \\ [+3pt] 
D$_2$O     &   1.7(-4)   &  $\leq$3.0(-4)              & \cite{coutens13} \\ [+3pt]
CH$_3$OD   &   2.0(-2)   &  1.8$^{+2.2}_{-1.2}$(-2)      & \cite{parise06} \\ [+3pt]
CH$_2$DOH  &   3.6(-1)   &  3.7$^{+3.8}_{-1.9}$(-1)      & \cite{parise06} \\ [+3pt]
CH$_2$DOD  &   7.2(-3)   &  -                          & - \\ [+3pt]
CD$_2$HOH  &   4.4(-2)   & 7.4$^{+8.4}_{-4.4}$(-2)       & \cite{parise06} \\ [+3pt]
CD$_2$HOD  &   8.8(-4)   &  -                          & - \\ [+3pt]
CD$_3$OH   &   1.8(-3)   & 8.0$^{+6.0}_{-6.0}$(-3)       & \cite{parise04} \\ [+3pt]
CD$_3$OD   &   3.5(-5)   &  -                          &  - \\ [+3pt]
\hline
\end{tabular}
\end{table*}

\subsection{High-mass protostar: Orion Compact Ridge}

\begin{figure}
\begin{center}
\includegraphics*[width=7.0cm,angle=-90.]{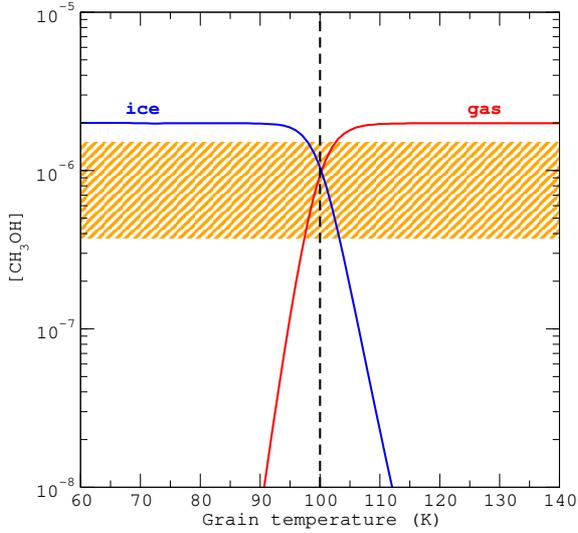}
  \caption{Gas phase and ice abundance of CH$_3$OH as a function of
    grain temperature. The abundance towards Orion~KL is represented
    by the orange hatched zone \citep{neill13}. The solid vertical
    dashed line denotes $T_d=100$~K. See text for details.}
\label{fig4}
\end{center}
\end{figure}

The steady-state gas- and ice-phase abundances (relative to $n_{\rm
  H}$) of CH$_3$OH in the Compact Ridge are plotted in Fig.~\ref{fig4}
as a function of dust temperature. In this plot, the initial water
fractionation ratio is again taken as [s-HDO]/[s-H$_2$O]=1\%. As
observed previously, the CH$_3$OH abundance increases sharply above
90~K and solid methanol is entirely sublimated for temperatures higher
than 110~K. The CH$_3$OH abundance in the Compact Ridge is estimated
as $\sim 7.5\times 10^{-7}$, with a column density of $6.0\times
10^{17}$~cm$^{-2}$ \citep{neill13}. The orange hatched zone in
Fig.~\ref{fig4} depicts the corresponding uncertainty of a factor of
2. It is found that the observed abundance is consistent with a dust
temperature of $\sim 100$~K. Thus, in contrast to IRAS~16293-2422, our
model suggests that a large fraction ($\gtrsim 40\%$) of the ice
mantles has  evaporated in the compact ridge (assuming that the
ice composition in the Compact Ridge is typical of high-mass
sources). In the following, a standard hot-core dust temperature of
100~K is employed.

\begin{figure}
\begin{center}
\includegraphics*[width=7.0cm,angle=-90.]{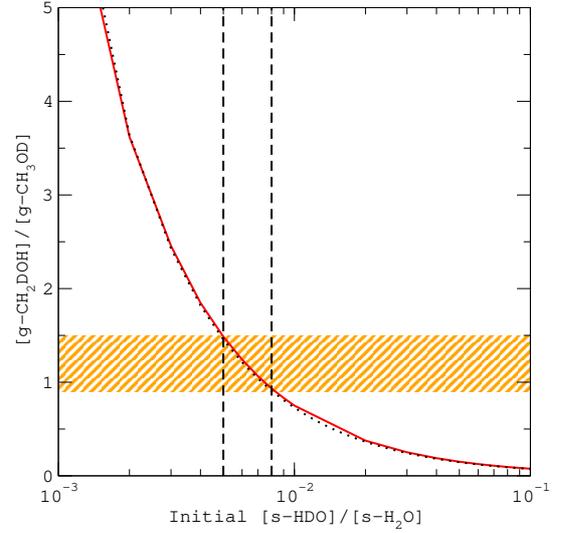}
  \caption{Gas-phase abundance ratio of the singly deuterated
    isotopologues of methanol, [g-CH$_2$DOH]/[g-CH$_3$OD], as function
    of the initial (cold) water ice deuteration. The dotted line
    corresponds to Eq.~(\ref{equi4}). The ratio towards Orion~KL is
    represented by the orange hatched zone \citep{neill13}.}
\label{fig5}
\end{center}
\end{figure}

The gas-phase ratio [g-CH$_2$DOH]/[g-CH$_3$OD] obtained with our model
is plotted in Fig.~\ref{fig5} as function of the initial water ice
deuteration [s-HDO]/[s-H$_2$O]. It is found to be below the
statistical value of 3, as soon as the initial water deuteration is
larger than $\sim 3\times 10^{-3}$. Thus, isotopic H/D exchanges
between methanol and water can also quantitatively explain  the small
[g-CH$_2$DOH]/[g-CH$_3$OD] ratios observed in high-mass protostars. In
particular, our model matches the observational ratio towards the
Compact Ridge ($\sim 1.2\pm 0.3$) for an initial [s-HDO]/[s-H$_2$O] in
the range $5-8\times 10^{-3}$. Again, the determined range of D/H
ratios for water ice is consistent with the upper limits derived
towards high-mass protostars, i.e. [s-HDO]/[s-H$_2$O]$<10^{-2}$
\citep{dartois03}. In contrast to
IRAS~16293-2422, the deduced primitive D/H ratio of water ice
([s-HDO]/[s-H$_2$O]$\sim 0.6\%$) is similar to that of methanol (see
Table~2)in that it corresponds to a statistical deuteration with an
accreting D/H atomic ratio $\alpha_m \sim \alpha_w\sim 2-3\times
10^{-3}$.

As above, we can employ Eq.~(\ref{equi2}) for a dust temperature
$T_d=100$~K and $\alpha_m=2\times 10^{-3}$, i.e.
\begin{equation}
\frac{[{\rm CH_2DOH}]}{[{\rm CH_3OD}]}=\frac{0.0073}{[{\rm
      HDO}]/[{\rm H_2O}]}
\label{equi4}
.\end{equation}
Equation~(\ref{equi4}), plotted in Fig.~\ref{fig5} as a dotted
line, reproduces very well the ratio [g-CH$_2$DOH]/[g-CH$_3$OD] for
all initial water deuterations over the range
$10^{-3}-10^{-1}$. Indeed, in this case the dominant reservoir of
exchangeable deuterium atoms in the ice is always water
([HDO]$>5\times 10^{-8}$ while [CH$_3$OD]=$4\times
10^{-9}$). Hydroxyl-deuterated methanol thus gets enriched in
deuterium through H/D exchanges with HDO, except when
[s-HDO]/[s-H$_2$O]$\lesssim 3\times 10^{-3}$ where the ratio
[g-CH$_2$DOH]/[g-CH$_3$OD] exceeds 3.

In Fig.~6 we plotted the D/H ratio of gaseous water as function of the
initial D/H ratio of water ice. As expected, the gas-phase ratio is
only (and slightly) increased with respect to the initial ice-phase
value when the initial [s-HDO]/[s-H$_2$O] is lower than $\sim
10^{-3}$. For the best value [s-HDO]/[s-H$_2$O]=0.6\%, the
[g-HDO]/[g-H$_2$O] ratio is thus found to be $6.0\times 10^{-3}$. This
is in good agreement with the observational values derived by
\cite{neill13} in the Compact Ridge,
     [g-HDO]/[g-H$_2$O]=$3.8^{+3.6}_{-2.5}\times 10^{-3}$. As a
     result, in this source, the post-evaporative chemistry of water
     seems negligible.

\begin{figure}
\begin{center}
\includegraphics*[width=7.0cm,angle=-90.]{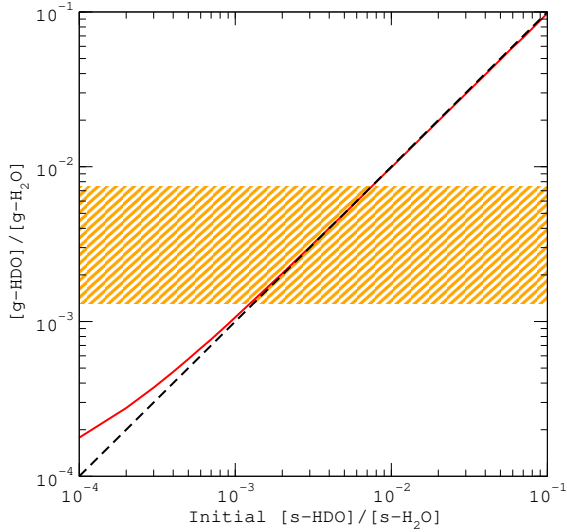}
  \caption{Gas-phase deuterium fractionation of water,
    [g-HDO]/[g-H$_2$O], as function of the initial water ice
    deuteration. The ratio observed towards Orion~KL is represented by
    the orange hatched zone \citep{neill13}.}
\label{fig6}
\end{center}
\end{figure}

Finally, in Table~4, we summarize the results of our best model,
corresponding to an initial [s-HDO]/[s-H$_2$O] of 0.6\%. A very good
agreement with the observations is obtained again, but we stress that
no multiply deuterated isotopologues are detected in the Compact
Ridge. Our model predicts in particular column densities for CH$_2$DOD
and CD$_2$HOH of $\sim 2.2\times 10^{13}$~cm$^{-2}$ and $\sim
9.2\times 10^{12}$~cm$^{-2}$, respectively, corresponding to
abundances (relative to $n_{\rm H}$) of $2.7\times 10^{-11}$ and
$1.2\times 10^{-11}$.

\begin{table*}[htp]
\centering
\caption{Comparison between observational fractionation ratios in the
  Orion Compact Ridge hot core source and our best model. Numbers in
  brackets denote the power of ten.}
\begin{tabular}{l c c c}
\hline
\hline
Species    & Best model  & Observations              &  References  \\ [+3pt]
\hline
HDO        &  6.0(-3)    &   3.8$^{+3.6}_{-2.5}$(-3)   & \cite{neill13} \\ [+3pt]
D$_2$O     &  1.0(-5)    &   -                        &  - \\ [+3pt]
CH$_3$OD   &  4.9(-3)    &   5.0$^{+1.0}_{-1.0}$(-3)    & \cite{neill13} \\ [+3pt]
CH$_2$DOH  &  6.0(-3)    &   5.8$^{+1.2}_{-1.2}$(-3)    & \cite{neill13} \\ [+3pt]
CH$_2$DOD  &  2.9(-5)    &   -                        & - \\ [+3pt]
CD$_2$HOH  &  1.2(-5)    &   -                        & - \\ [+3pt]
CD$_2$HOD  &  2.5(-8)    &   -                        & - \\ [+3pt]
CD$_3$OH   &  8.2(-9)    &   -                        & - \\ [+3pt]
CD$_3$OD   &  1.7(-11)   &   -                        & - \\ [+3pt]
\hline
\end{tabular}
\end{table*}

\section{Discussion}

The above chemical kinetics model can be summarized by the following
three phases: first, the {\it \textup{cold accretion phase}}, corresponding to
the prestellar stage of star formation. In this initial phase, we have
assumed that the D/H ratio of both water and methanol is statistical
and proceeds in the ice through the addition of H and D atoms on solid
O and CO, respectively. This deuteration process takes place when the
dust temperature is around 10~K and water ice is amorphous. The key
parameter during this phase is the accreting D/H gas-phase ratio,
which controls the level of deuteration in the ice. This parameter was
fixed for methanol by assuming that the fractionation of the gas-phase
CH$_2$DOH, as observed in hot cores, reflects the primitive methanol
ice deuteration. As the accreting D/H ratio increases with time (see
below), as well as the O and CO depletion onto grains, a different
fractionation for water was explored. At this stage, the ratio
[s-CH$_2$DOH]/[s-CH$_3$OD] is equal to 3, which is the statistical value. In
the second stage, {\it \textup{the warm-up phase}}, the dust mantles are heated
by the nascent protostar. In our model, the temperature jumps from
10~K to 100~K are instantaneous. During this phase, amorphous water is
converted to the stable cubic crystalline form and this reorganization
process is accompanied, in its very initial phase, by the mobility of
water and the lability of protons and deuterons. Hydrogen isotope
exchanges are triggered during this phase, as demonstrated
experimentally by \cite{faure15}. At this stage, the ratio
[s-CH$_2$DOH]/[s-CH$_3$OD] is no longer equal to 3 because the -OH
groups of methanol have equilibrated with water ice. The
[s-CH$_2$DOH]/[s-CH$_3$OD] ratio is therefore inversely proportional
to the [s-HDO]/[s-H$_2$O] ratio. In the third stage, {\it \textup{the hot core
  phase}}, the mantles are (partially or entirely) sublimated and the
[g-CH$_2$DOH]/[g-CH$_3$OD] ratio established in the ice during the
warm-up phase is transferred (and conserved) into the gas phase.

The main result of this work is that except for the -CH functional
groups, the gas-phase molecular D/H ratios measured in hot cores
cannot be employed to infer the primitive deuteration in the
ices. Molecules with functional groups able to establish hydrogen
bonds (e.g. -OH or -NH) are indeed expected to equilibrate with water
ice during the warm-up phase, with short timescales with respect to
desorption. For these species, the fractionation ratio [XD]/[XH]
should scale inversely with the fractionation of water through the
equilibrium constant. We have shown that this is the case for
hydroxyl-deuterated methanol and that the [g-CH$_2$DOH]/[g-CH$_3$OD]
ratios measured in hot cores reflect H/D equilibrium in water ice:
ratios above the statistical value of 3 show evidence of a large
fractionation of methanol relative to water, while values below 3
attest a similar fractionation for methanol and water. As a result,
the [g-CH$_2$DOH]/[g-CH$_3$OD] ratio is a very sensitive probe of the
[s-HDO]/[s-H$_2$O] ratio, for which only upper limits can be
determined observationally \citep{galvez11}. This result certainly
extends to other interstellar molecules containing different
functional groups, such as formic acid (HCOOH), methylamine
(CH$_3$NH$_2$), formamide (HCONH$_2$), etc. We emphasize that the H/D
exchange rate constant between methylamine and water has been measured
in the laboratory by \cite{faure15}. The activation energy was found
to be even lower ($\sim 3300$~K) than in the case of methanol. To the
best of our knowledge, however, the deuterated isotopologues of
CH$_3$NH$_2$ have not been detected yet in the ISM. Another
interesting candidate is ammonia (NH$_3$), which is abundant in ice
($\sim 5$\% with respect to water) and for which all
multiply-deuterated isotopologues have been identified
\citep{roueff05}. H/D exchanges between NH$_3$ and D$_2$O have been
observed recently by \cite{lamberts15}. H/D exchanges between water
and -NH bearing molecules will be adressed in a future, dedicated
study. In contrast, molecules containing only -CH functional groups,
such as H$_2$CO, should conserve their primitive fractionation. The
particular case of HCN was also investigated by \cite{faure15} and no
proton exchange was observed.

We stress that the chemical processes involved in our kinetics model
are identical in low- and high-mass sources, and the main difference
lies in the initial fractionation ratios. By comparing our model with
hot core observations, we have thus found that the primitive
deuteration of water and methanol ices was similar in the Orion
Compact Ridge ($\alpha_m \sim \alpha_w \sim 0.2-0.3$\%) while the
deuteration of methanol ices was much higher than that of water in
IRAS~16293-2422 ($\alpha_m\sim$12\% and $\alpha_w\sim$1\%). Our
results therefore suggest that the molecular D/H ratio in the ice
mantles was rather homogeneous in Orion but highly heterogeneous in
IRAS~16293-2422. This result may be linked to different dynamical
timescales of the prestellar core phases, the longer timescales in
low-mass sources allowing a larger gas-phase [D]/[H] ratio. It is in
particular striking that the level of water deuteration is similar in
both sources (i.e. about 1\%), while the methanol deuteration is larger
by an order of magnitude in IRAS~16293-2422. The differential
deuteration of the grain mantles in low-mass protostars was modelled
recently by several authors
\citep{cazaux11,taquet12,taquet13,taquet14}. In the multi-layer
approach of Taquet et al., in particular, an increase of the D/H ratio
towards the mantle surface is predicted. This result is a direct
consequence of the increase of the accreting D/H ratio with time. In
these models, the CO depletion factor and the ortho-to-para (OPR) ratio
of H$_2$ are of crucial importance because they control the abundance
of the deuterated isotopologues of H$_3^+$, which are the main
deuteron donors in the cold gas. The D/H ratios in the ice mantles can
therefore vary, from prestellar core to prestellar core for a given
species, and from species to species for a given core. In agreement
with the results of \cite{taquet14}, our model indicates that
differential deuteration occured in the ices of IRAS~16293-2422, where
the bulk of water ice was formed at low accreting D/H ratio
(i.e. early times), while the bulk of methanol ices was formed at later
times in a gas enriched in atomic deuterium. This does not exclude a
high D/H ratio for water in the uppermost layers of the grains: these
HDO molecules  also equilibrate with the water bulk upon
heating. In the presence of ice segregation,
however, H/D exchanges might be limited because some methanol
molecules may not be surrounded by water before sublimating. This
might occur in ices where the methanol abundance is much higher than a
few percent \citep[see e.g.][]{dartois99} in which case the gas-phase
[CH$_2$DOH]/[CH$_3$OD] ratio could be close to statistical.

Finally, hydrogen isotope exchanges are expected to play an important
role in astronomical environments other than hot cores. In the
interstellar medium, whatever the desorption mechanism, i.e. thermal
heating, cosmic-ray impacts or UV photon induced desorption,
accompanying changes in the structure of water ice are expected to
promote H/D exchanges. In this context, it is worth  mentioning the
detection of CH$_3$OH and CH$_2$DOH in prestellar cores where the dust
temperature is $\sim 10$~K \citep{bacmann07,bizzocchi14}. In these sources, the presence of gas-phase methanol demonstrates the occurence
of non-thermal (or locally thermal) evaporation processes. The
derived [CH$_2$DOH]/[CH$_3$OH] fractionation ratios were found in the
range 5-30\%, i.e. slightly smaller than the values measured in hot
corinos, with upper limits on the column density of CH$_3$OD
corresponding to [CH$_2$DOH]/[CH$_3$OD]$\geq 10$, i.e. much larger
than the statistical value \citep{bacmann07,bizzocchi14}. This result
may suggest the presence of H/D exchanges triggered by the non-thermal
desorption. Another obvious application of H/D exchanges is in
comets. Any molecules containing -ND or -OD groups in a cometary
nucleus should indeed experience thermal H/D exchanges with water when
the comet approaches the Sun. This was in fact postulated by
\cite{blake99} to explain correlated measurements between
     [HDO]/[H$_2$O] and [DCN]/[HCN] ratios in comet Hale-Bopp. H/D
     exchanges between HCN and water are however not supported
     experimentally \citep{faure15}. To our knowledge, no other
     deuterated molecules have been detected so far in comets. Results
     from the ROSETTA mission towards comet 67P/Churyumov-Gerasimenko
     should provide new clues in this context \citep{altwegg15}.

\section{Conclusions}

We have reported a simplified kinetics model of the deuterium
chemistry occurring in the icy mantles of interstellar grains. The
chemistry was based on a statistical initial deuteration, i.e. the
statistical addition of H and D atoms on cold O and CO, followed by an
instantaneous warm-up phase during which thermal H/D exchanges occur
between water and methanol. The kinetic rate constants for H/D
exchanges were taken from laboratory data. The model was applied to
the chemistry of hot corinos and hot cores, with IRAS~16293-2422 and
the Orion~KL Compact Ridge as prototypes, respectively. For the first
time, our model was able to reproduce quantitatively the observed
gas-phase [CH$_2$DOH]/[CH$_3$OD] ratios, which are found to scale
inversely with [HDO]/[H$_2$O] owing to the H/D exchange equilibrium
between the hydroxyl (-OH) functional groups of methanol and water in
the ice. Primitive fractionations of water ice [HDO]/[H$_2$O]$\sim$2\%
in IRAS~16293-2422 and $\sim 0.6$\% in the Compact Ridge were deduced,
suggesting highly hetereogeneous D/H ratios in the ices of
IRAS~16293-2422. Molecules with functional groups able to establish
hydrogen bonds (e.g. -OH or -NH) are thus expected to equilibrate with
water ice during any thermal (or non-thermal) heating event. Similar
results are expected in other astronomical environments, in
particular, in cometary nuclei. This shows the relevance of using a
reduced microphysics code based on laboratory experiments to interpret
astronomical observations. We therefore strongly recommend
incorporating the H/D exchange kinetic data in any detailed model of
the deuterium chemistry in interstellar or cometary ices.


\begin{acknowledgements}

This work was supported by the CNRS program ``Physique et Chimie du
Milieu Interstellaire'' (PCMI). One of us (M.F.) is also supported by
a fellowship of the Minist\`ere de l'Enseignement Sup\'erieur et de la
Recherche. Aurore Bacmann, Pierre Hily-Blant and S\'ebastien Maret are
acknowledged for useful discussions.

\end{acknowledgements}
  
\bibliographystyle{aa}
\bibliography{biblio}

\end{document}